# Cathodic arc voltage noise modelling: ectons and random walk of the cathode spots

Arshat Urazbayev

*National Laboratory Astana, Laboratory of Computer Sciences, Nazarbayev University, 010000 Astana, Kazakhstan*

*Abstract.* The noise signal of the voltage fluctuations of cathode arcs recorded at different sampling frequencies was analyzed. To explain obtained Fourier spectra for each measurement, a model was developed. The characteristics of the model were obtained by comparing Fourier spectra of the developed model and experimental signals. The model shows good quality agreement with the experiment. The voltage signal consists of two parts: a sum of very short (several ns) fluctuations, related a microexplosion or ecton and a low frequency part related to a random cathode spot walk.

Cathode arc spot was intensively studied by different authors [1, 2, 3]. The mechanisms of these processes are still not clear. Previously, A. Andres et al. studied the noise of the burning voltage of cathodic arcs in vacuum for various cathode arcs [4]. The arcs were generated in a coaxial plasma source and fluctuations of voltage during arc were recorded using a broadband measuring system. In this work, the model of the arc noise based on previous experimental data is developed.

Fig. 1 shows the schematic set-up of the experiment [4]. Coaxial plasma source was used and to preserve the frequency characteristics of the voltage noise, the arc source was designed like a broadband line, e.g., by choosing a coaxial design. The center cathode was a rod of 6.25 mm diameter surrounded by a coaxial anode cylinder. On one end, the plasma source was essentially comprised of a high-current vacuum feedthrough. In [4] the parameters of the experiment were as following: the arc was fed by a large, 0.33 F capacitor bank and switched with a high-voltage and high-current transistor. For most of the experiments, the arc had a current of 100 A for a duration of 2.5 ms. All arc experiments were done in vacuum at a base pressure of about $10^{-4}$ Pa without any process gas. The arc burning voltage was monitored directly at the vacuum feedthrough using a broadband 100:1 divider Tektronix P5100, max. 250 MHz. Voltage data were captured in sample mode 50 000 data points per arc with various time resolutions and their associated sample rates. The data were then exported from the scope for a further analysis.

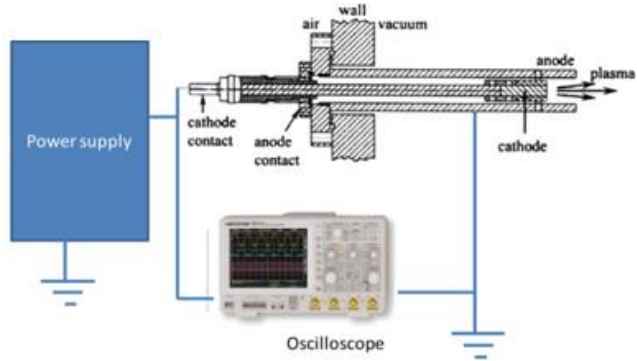

Fig. 1. Schematic of the experimental set-up with a coaxial plasma source.

Here, series of experiments with molybdenum cathode at different sampling frequency 5MHz, 20MHz and 100MHz were used. At every sampling frequency 50,000 points of array data was collected from the voltage versus time signal during the operation of the cathode arc. Obtained data upon zooming in time scale shows a high level of the quantization due to one-tenth precision after the decimal point in the voltage measurements.

Fourier spectra of molybdenum arc with three different sampling frequencies are shown in Fig.2. In the spectrum with sampling frequency of 100MHz, there are two parts: a linear slop at the frequencies from $10*10^6$ to $5*10^7$ Hz and curved part at $2*10^8$ Hz (Fig. 2A). Quantization noise at $2*10^8$ Hz and higher was not used for the further analysis. Low frequency part is observed more clearly in the 20MHz spectrum (Fig. 1B). In the spectrum of 5 MHz, a linear part is prolonged, while a curved part is not pronounced, suggesting that this feature appears only when a degree of the quantization noise is very high (Fig. 2C).

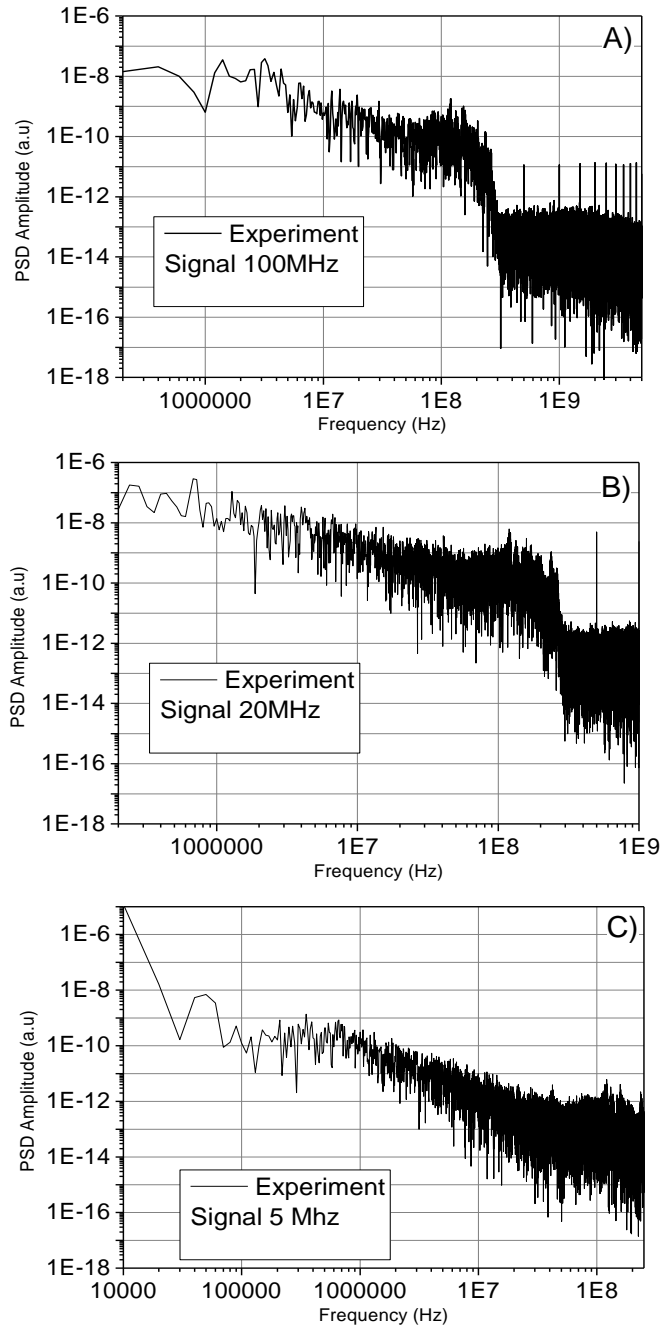

Fig.2. Fourier spectra of the experimental signal at different frequencies. A) 100 MHz, B) 20 MHz, C) 5 MHz.

## *Developing the model of the noise*

Here, the proposed model was based on the study of the tokamak turbulence [5][6]. Observed two parts of the spectra are related to the different types of fluctuations. First, curved part at the higher frequencies is considered. The arc consists of small individual explosions. Each explosion produces the burst of electrons and plasma. According to Mesyats et al. [7], these bursts of the explosive release of electrons and plasma cannot be arbitrarily small but need to exceed some critical amount called "ecton". Microscopic craters on the cathode surface are the evidence for these micro explosions. They facilitate an emission of a limited charge with an average value defined by $< Q > = < \int_0^{\partial t} I dt >$, where $<>$ brackets are averaging over a large number of the emission events and $\partial t$ indicates the duration of the emissive phase of a spot cell also known as a "lifetime". The goal is to find $\partial t$ from the experimental signal. Measured voltage versus the time was modeled as a sum of individual time-depended function $u(t)$, which defines each fluctuation that happens sequentially in the form of micro explosions on the cathode surface during the arc operation (Fig.3). The following formula was used as a model:

$$U_e(t) = \sum_{k=1}^{n} u(t - \tau_k) \quad (1)$$

where $U_e$ (e stands for ectons) is a function of the voltage related to ectons, n is a total number of fluctuations, $\tau_k$ is a random value that represents the inception time of $k^{th}$ fluctuation and $u(t)$ is a time-depended function of the voltage for each fluctuation. It is assumed, that the function is the same for all fluctuations and $\tau_k$ is uniformly distributed. Total number of the discharges should be high enough to get the mean value of $U_e(t) >> max(u(t))$ implying that every unit function $u(t)$ should overlap. Next, the unit function $u(t)$ is defined. The function was approximated using modified Loretzian function:

$$u(t, \tau_k, d_{sim}) = A \left( \frac{1}{1+\left(\frac{x-\tau_k}{d_{sim}}\right)^3} - 0.1 \right) \quad (2)$$

where if $u(t, \tau_k, d_{sim}) < 0$ then $(t, \tau_k, d_{sim}) = 0$ making the function confined in time, since the microexplosions should be confined in time. A is an amplitude of the fluctuation and $d_{sim}$ is a characteristics time scale of the fluctuation, which can be approximated as the duration of the microexplosion $\partial t = 2 * d_{sim}$. By changing $d_{sim}$, variations of $U_e(t)$ function are obtained. Then Fourier spectrum of the model signal $U_e(t)$ and experimental were compared. It was shown that the Fourier spectrum had no dependence on a number of the fluctuations. Even few fluctuations had the same average spectrum as a sum of thousand ones. Here, reasonably large number of the fluctuations were used, where average value of the simulated signal was several times more than the amplitude of one fluctuation. Therefore, to simplify the analysis a constant part of the signal was deducted. Additionally, both signals, model and experiment, were normalized at $1*10^8$ Hz. The fitting of the model signal was also adjusted by adding the white noise to simulate the quantization noise of the experimental signal.

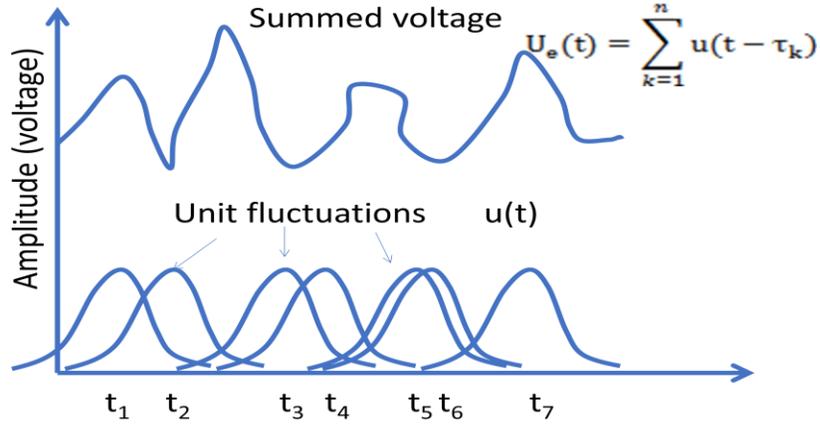

Fig. 3. Simulation of the voltage signal. Voltage signal is a sum of small unit fluctuations.

Fig.4 shows that model signal is in a good agreement with experimental signal. Both spectra have maximum at zero frequency and feature curves at $2*10^8$ Hz. Parameter $d_{sim}$ is equal to 1.5 ns and a mean time of life for ectons $\partial t = 3$ ns. The model with a constant value for $d_{sim}$ showed a better fitting than the model with was varied $d_{sim}$, suggesting the ecton origin of these fluctuations.

There is a characteristic curve at about $2.3*10^8$ Hz and the frequency of this characteristic curve depends on $d_{sim}$. Therefore, the lifetime of ecton can be written as:

$$\partial t = 2 * d_{sim} = C * \frac{1}{f_c} \tag{3}$$

where $f_c$ is a frequency of the curve and C is a constant. This constant is different for different functions $u(t, \tau_k, d_{sim})$, in this case it equals to 1.4, while $\frac{1}{f_c}$ equals to wavelength of the mode. Using (3) lifetime of ecton can be easily estimated if there is a characteristic curve in the spectrum.

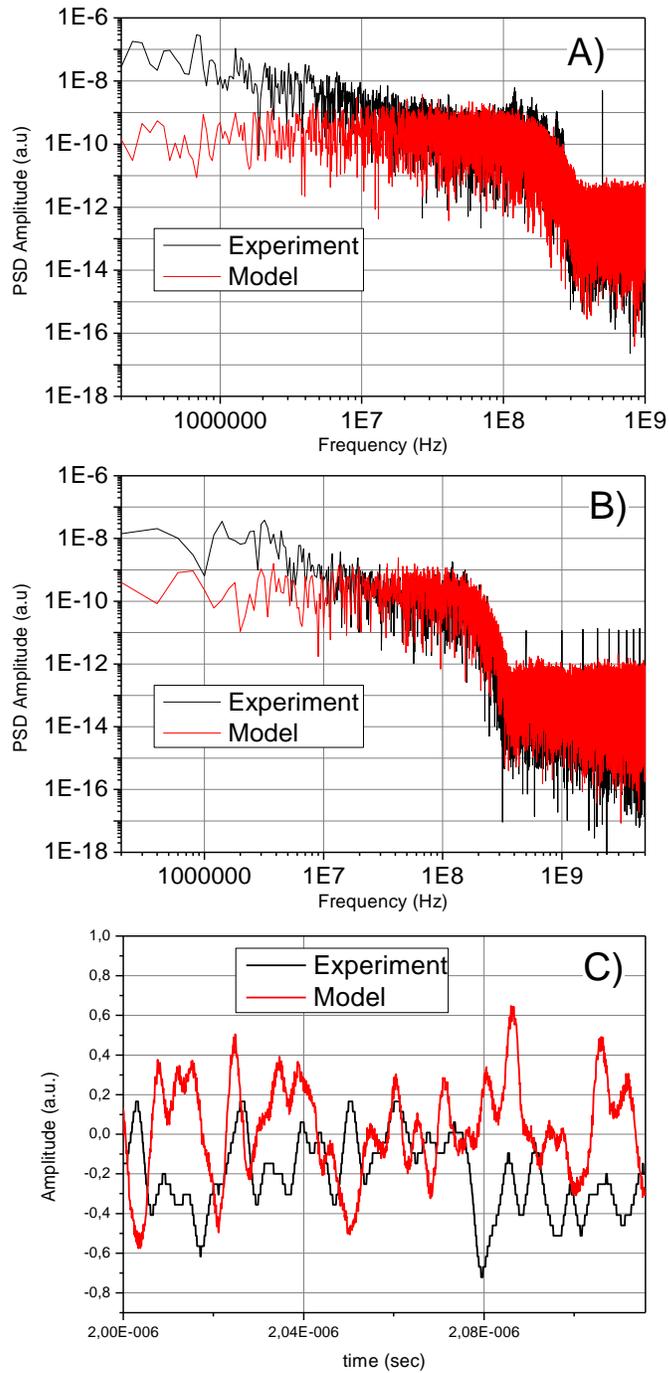

Fig. 4. Comparison of Fourier spectra of experimental end modeling signals at high frequency fluctuations for sampling frequencies: A) 20MHz, B) 100MHz. C) Normalized experimental (black), and model signals (red).

Next, low frequency linear part of the spectra is analyzed. In [4], it was suggested that the linear part of spectra related to the Brownian noise and this type of noise attributed to the random walk of the cathode spot [8,9,10]. Here, it is assumed that the oscillations responsible for the linear spectrum fluctuations were the local fluctuations of the ectons number, suggesting that the minimum of a low frequency part of the signal relates to the minimum concentration of ectons, and the maximum relates to the maximum concentration. According to [10] the trace of arc has the tree–like (or fractal–like) structure. In the current model, it is assumed that low frequency fluctuations are to some extent consequences of this structure. Therefore, every brunch of this tree structure can be described as a sequence of microexplosions or ectons with the limited duration or lifetime. Additionally, every brunch can produce another brunch with the limited lifetime and it is a random process. In this process, there is a constant probability of the branch's birth in a sampling time. Therefore, the low frequency part of spectra relates to the oscillation of a brunch number in this moment due to the difference between the birth and death of brunches.

Every brunch is confined in a time function $U_e(t)$. The voltage of the brunch can be calculated from (1). Lifetime $t_{lifetime}$ is a random number between the values of $t_{timelifemax}$ and $t_{timelifemin}$. Once one-time step is made, current time is added to the sampling time and a new cluster with some probability is also added. If the age of the brunch exceeds its $t_{lifetime}$, it is removed from the tree. At any given moment, the voltage of the system will be a sum of all amplitudes of all brunches in the system:

$$U_{rw}(t) = \sum_{k=1}^{n(t)} U_e^k(t) \qquad (4)$$

where $U_{rw}$ (rw stands for a random walk) is a function of the voltage related to the random spot walk, n(t) is a total number of the brunches. Fig. 5 shows the simplified scheme of the model. At $t_1$ the number of brunches is three, at $t_2$ there is only one and $t_3$ there are two brunches. Therefore, at $t_1$ $U_{rw}(t)$ is a local maximum and at $t_2$ local minimum. Total number of brunches at any given time varied from forty up to several hundreds.

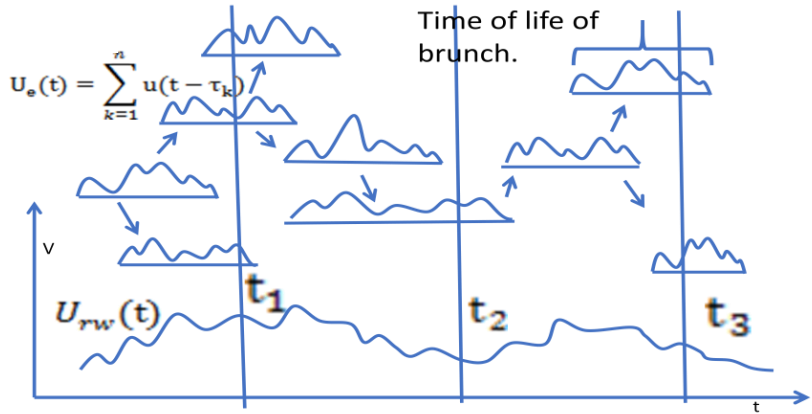

Fig. 5. Tree-like structure of the arc trace. The scheme of a model of the signal.

The model for the sampling frequency 5 MHz is in a good agreement with experimental signal (Fig. 6). Parameters used for the model are $t_{imemin}$ = 15 ns, $t_{timrmax}$ = 900 ns, the birth probability = 10% per 2 ns. For the low frequency linear part for 5 MHz sampling frequency $t_{lifetime}$ is a variable parameter with a microsecond range, which corresponds to the high variations of the branch lengths. The characteristic curving at $2.3*10^8$ Hz is also clearly seen on the model spectra.

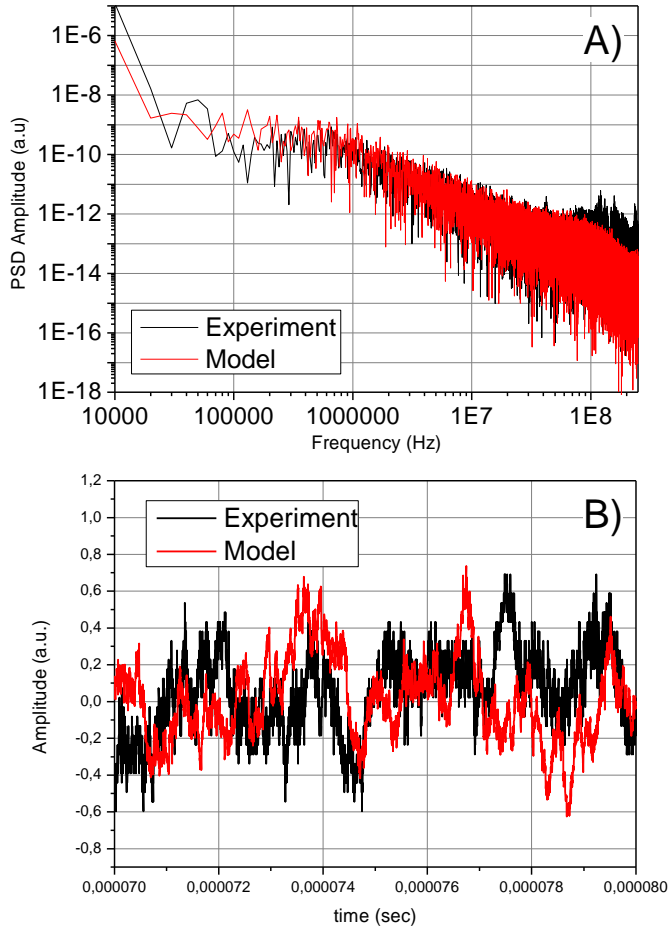

Fig. 6. Comparison of Fourier spectra of the experimental and model signals at the low frequency linear part fluctuations for the sampling frequency A) 5MHz. B) Normalized experimental (black) and model (red) signals.

During the model development, a constant value of $t_{lifetime}$ was also tested (Fig.7). However, maximum peak appeared in the model spectra, while in Fig.2, no maximum was observed for high frequency part of the spectrum, where $d_{sim}$ is constant. Therefore, $t_{lifetime}$ value should vary in the range of up to microseconds. It can be explained by the fact that the length of the brunches varies a lot as well.

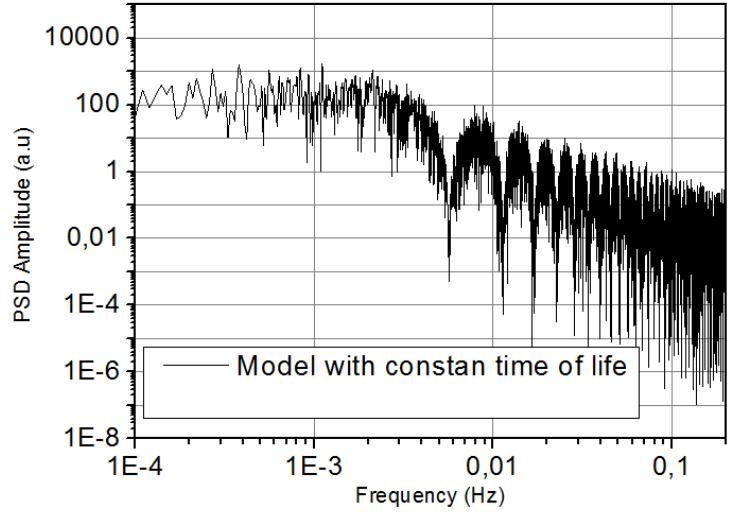

Fig 7. Fourier spectra of the model signals at the low frequency linear part fluctuations lifetime of model.

In [4] Anders et al. performed experiments for different cathode materials such W, Mg, Ti and others. Unfortunately, for most of the cathode materials no data with a sampling frequency higher than 20MGhz is available. Since the time of the microexplosion is about several nanoseconds and the main feature of the microexplosion in the spectrum (curve at $2*10^8$ Hz) is a white noise part of the spectrum, it is impossible to make a full-scale modeling for other cathode materials.

Here, the raw approximation of a spectrum for tungsten cathode is performed. It has a maximum cohesive energy among all of the cathode materials and it has the biggest difference in the Fourier spectrum comparing to the noise spectrum of molybdenum cathode. Model function is taken as a function of number of brunches at time t.

$$G(t) = \sum_{k=1}^{n(t)} U_k(t) = n(t) \qquad (5)$$

Fig 8 shows the fitting of a function of brunch number and experimental signal of the tungsten cathode arc. Model parameters are following: $t_{timemin} = 15$ ns, $t_{timamax} = 900$ ns and the probability of the birth = 50% per 2 nanoseconds (same time frequency as in experiment – 5Mhz) compared to 10 % for molybdenum. The average number of brunches for molybdenum is about 20, for tungsten is about 90.

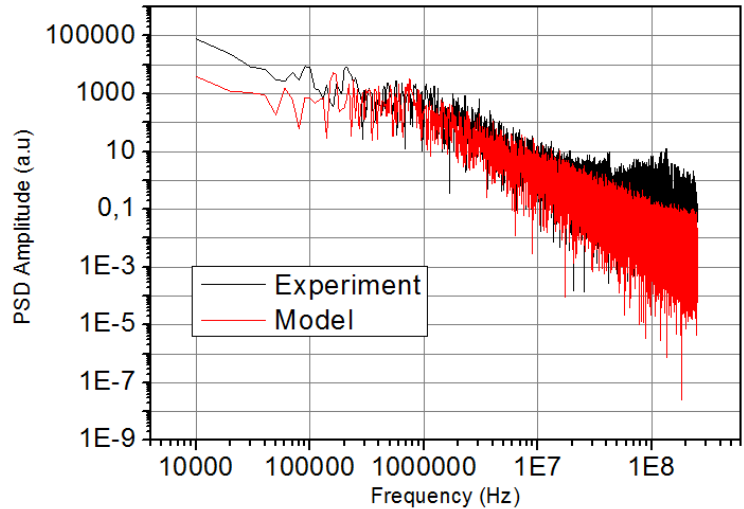

Fig.8. The comparison of the Fourier spectra of the experimental end modeling signals for low frequency linear part fluctuations for W.

*Conclusion.*

The model shows a good agreement with the experimental signal. The voltage signal consists of two parts: sum of very short (several ns) fluctuations, named microexplosion or ecton. This part relates to the high frequency part of Fourier spectrum. While the low frequency part (linear part of the spectrum) relates to the different phenomena – the random cathode spot walk and tree-like structure of the arc. The arc consists of brunches, and the lifetime of any brunch can vary in the range of microseconds. The birth probability of a new brunch is 10% per 2 ns.

*Acknowledgements*: The author would like to thank Andre Anders for the provided raw data and discussion of the results and Dana Akilbekova for the help with a discussion part. This work was supported by Ministry of Science and Education of The Republic of Kazakhstan (Program "NU-LBNL" 0115PK03029).